\documentclass{moriond}

\usepackage{geometry,amsmath,amsfonts,amssymb}
\usepackage{slashed}
\usepackage{subfig}

\def\be{\begin{equation}}
\def\ee{\end{equation}}
\def\bea{\begin{eqnarray}}
\def\eea{\end{eqnarray}}

\newcommand{\Photo}{\includegraphics[height=35mm]{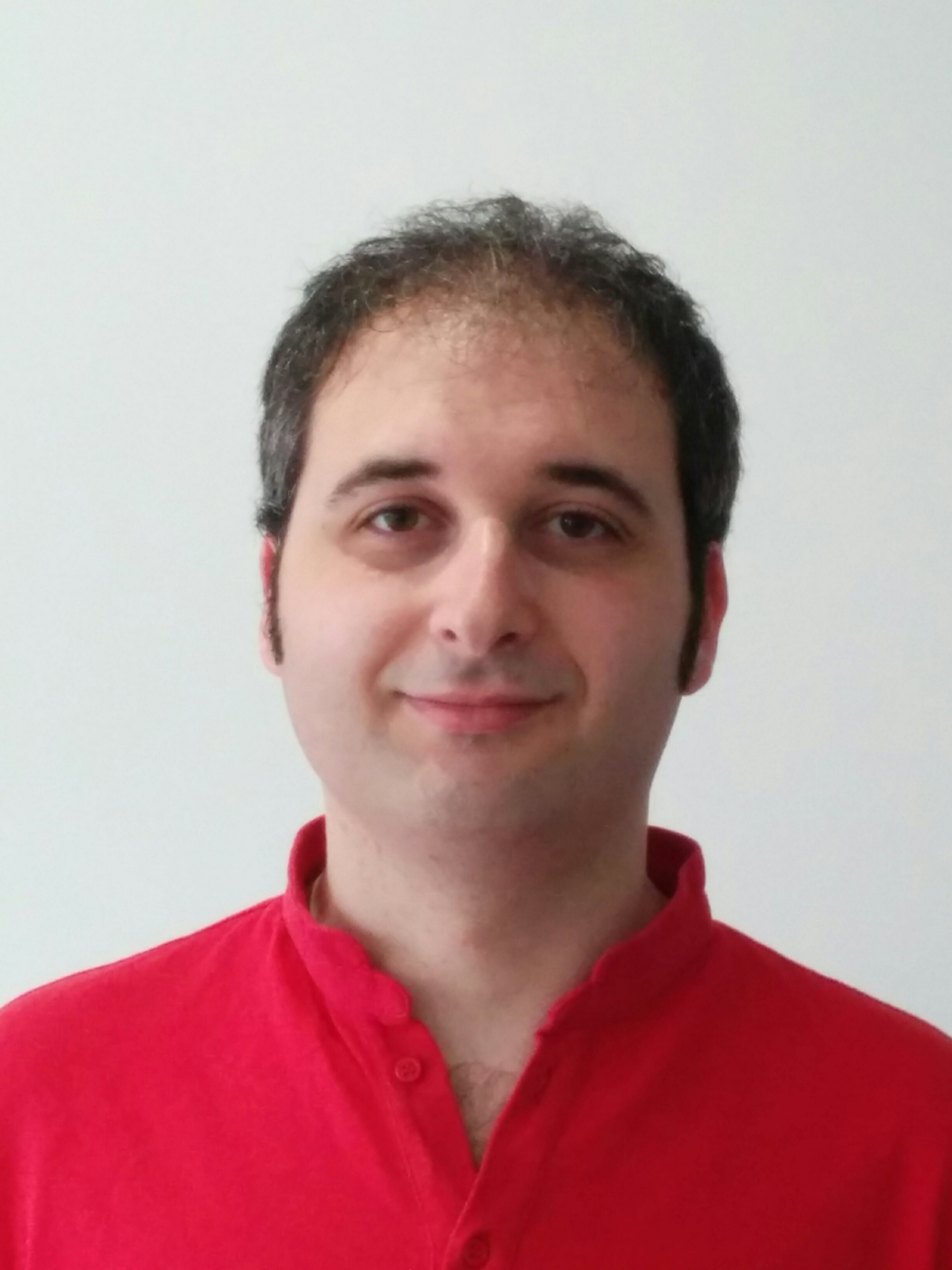}}
\newcommand{\bee}[0]{\begin{eqnarray}}
\newcommand{\eee}[0]{\end{eqnarray}}

\begin{document}
\vspace*{4cm}
\title{LEPTON NUMBER SYMMETRY AS A WAY TO TESTABLE LEPTOGENESIS}

\author{M. LUCENTE${}^1$, A. ABADA${}^2$, G. ARCADI${}^2$ AND V. DOMCKE${}^3$}

\address{${}^1$Centre for Cosmology, Particle Physics and Phenomenology (CP3)\\
Universit\'e catholique de Louvain,
Chemin du Cyclotron 2, B-1348 Louvain-la-Neuve, Belgium\\
${}^2$Laboratoire de Physique Th\'eorique, CNRS -- UMR 8627, \\
Universit\'e de Paris-Sud, F-91405 Orsay Cedex, France\\
${}^3$AstroParticule et Cosmologie (APC)/Paris Centre for Cosmological Physics\\
Universit\'e Paris Diderot, Paris, France}

\maketitle\abstracts{We propose a minimal and motivated extension of the Standard Model characterised by an approximate lepton number conservation, which is able to simultaneously generate neutrino masses and to account for a successful baryogenesis via leptogenesis. The sterile fermions involved in the leptogenesis process have masses at the GeV scale. 
We determine the viable parameter space that complies with both the neutrino and baryogenesis phenomenology, and analyse the different regimes for the generation of a lepton asymmetry in the early Universe (weak and strong-washout) in order to determine their testability in future experimental facilities.}

\section{Introduction}
The measured baryon asymmetry of the Universe (BAU), $Y_{\Delta B} = (8.6 \pm 0.01) \times 10^{-11}$, is a robust observation that calls for the existence of physics beyond the Standard Model (SM): although a net baryon asymmetry can be produced within the SM framework in the hypothesis of a first order electroweak phase transition, it has been shown that the amount of CP violation in the quark sector is too small to account for the observed value~\cite{Gavela:1993ts,Huet:1994jb}. Baryogenesis in the SM is offered by sphalerons, which are non-perturbative configurations of the Higgs and electroweak gauge boson fields that mediate transitions violating both baryon and lepton number, while preserving their difference $B-L$. Sphalerons are in thermal equilibrium for a large range of temperature during the evolution of the Universe (${130 \text{ GeV} \lesssim T \lesssim 10^{12} \text{ GeV}}$), thus if any lepton asymmetry is produced during this epoch, the sphalerons convert it into a net baryon asymmetry: this is the so called baryogenesis via leptogenesis mechanism.

On the other hand there is another firm observation that calls for the existence of physics beyond the SM: the fact that neutrinos are massive and mix. It is remarkable that one of the most minimal extensions of the SM conceived to account for this observation, which is the Type-I seesaw mechanism, automatically provides the ingredients for a successful leptogenesis too. The Lagrangian of the Type-I seesaw reads
\bee\label{eq:lag_ssI}
\mathcal{L} &=& \mathcal{L}_\text{SM} + i \overline{N_I} \slashed{\partial} N_I - \left( Y_{\alpha I} \overline{\ell_\alpha} \tilde{\phi} N_I + \frac{M_{IJ}}{2} \overline{N_I^c} N_J + h.c.\right),
\eee
where $\ell_\alpha$ are the SM lepton doublets, $\phi$ is the Higgs doublet, $N_I$ are new fermionic fields that are singlets under the SM gauge group, $Y_{\alpha I}$ are dimensionless Yukawa couplings and $M$ is a matrix of Majorana mass terms for the $N_I$ fermions. After the electroweak symmetry breaking the Higgs field acquires a non-vanishing vacuum expectation value $v$, and the Lagrangian~(\ref{eq:lag_ssI}) accounts for a non-vanishing neutrino mass matrix $m_\nu$ which, under the assumption $v |Y_{\alpha I}| \ll |M_{IJ}|$, is given by
\bee\label{eq:numSS}
m_\nu \simeq - \frac{v^2}{2} Y^* \frac{1}{M} Y^\dagger.
\eee
As anticipated the Lagrangian~(\ref{eq:lag_ssI}) provides the necessary ingredients for a successful baryogenesis: the Yukawa couplings are in general complex numbers, providing an additional source of $CP$ violation; the lepton number violation of the physical neutrinos is converted to a baryon number violation via sphalerons, while the new sterile fermions can deviate from thermal equilibrium during their evolution. 

\section{Two different leptogenesis realisations}
The Lagrangian~(\ref{eq:lag_ssI}) is at the basis of one of the most popular realisations of the leptogenesis mechanism: thermal leptogenesis~\cite{Fukugita:1986hr}. In this mechanism it is assumed that the Yukawa couplings are sufficiently large such that the new sterile states are in thermal equilibrium in the very early Universe, $|Y_{\alpha I}| \gtrsim Y_{eq} \simeq \sqrt{2}\times 10^{-7}$. However, due to the Universe expansion, these states eventually feature a decoupling and a subsequent decay out of thermal equilibrium. Being Majorana particles their out of equilibrium decay results in the production of a net lepton asymmetry. If the lepton asymmetry is created before the electroweak phase transition the sphalerons convert it into a net baryon asymmetry. The thermal leptogenesis mechanism is very attractive from a theoretical point of view, but is prohibitive to test in laboratory experiments: in order to reproduce the observed amount of baryon asymmetry the mass scale $M$ of the new sterile fermions must be quite large~\cite{Abada:2006ea,Davidson:2008bu}, $M > 10^8$ GeV. This lower bound on the mass scale can be relaxed up to the TeV scale in the case of a mass spectrum containing pairs of sterile fermions that are strongly degenerate in mass~\cite{Pilaftsis:2003gt} (resonant leptogenesis), or down to $\mathcal{O}$(100 GeV) if in addition motivated flavour patterns are present~\cite{RLl}.

There is however a different mechanism to realise leptogenesis from the Lagrangian~(\ref{eq:lag_ssI}), known as flavoured leptogenesis~\cite{Akhmedov:1998qx} (or ARS mechanism). In this scenario it is assumed that the Yukawa couplings $Y_{\alpha I}$ are sufficiently small such that the new sterile fermions $N_I$ are out of thermal equilibrium in the early universe. If they do not equilibrate before the electroweak phase transition they provide the necessary deviation from thermal equilibrium. Rewriting eq.~(\ref{eq:numSS}) as
\bee
m_\nu \simeq - \frac{v^2}{2} Y^* \frac{1}{M} Y^\dagger \simeq 0.3 \left( \frac{\text{GeV}}{M} \right) \left( \frac{Y^2}{10^{-14}}\right) \text{ eV},
\eee
we can infer that in order to account for the observed values of neutrino masses and having Yukawa couplings sufficiently small such that the new sterile fermions do not equilibrate, the mass scale $M$ must lie around the GeV scale, opening the possibility of directly producing the new states in laboratory experiments. Notice however that this mass scale is much smaller than the temperature at which the leptogenesis process takes place, $M \sim \text{GeV} \ll T$, with ${T > 140\mbox{ GeV}}$. Thus the masses of the new sterile fermions, as well as their Majorana character, can be safely neglected and the total lepton number is conserved by the new interactions assumed in~(\ref{eq:lag_ssI}). The generation of the baryon asymmetry in the flavoured leptogenesis scenario proceeds in a different way with respect to the thermal case~\cite{Akhmedov:1998qx,ref:flavoured,Abada:2015rta}: in a first phase the sterile fermions are created from the interactions with the thermal plasma, the most important one being the scattering with the top quarks mediated by the Higgs boson. Then the Yukawa couplings in eq.~(\ref{eq:lag_ssI}) can cause a sterile fermion $N_I$ to oscillate into a different flavour $N_J$, $I\neq J$, via a non-diagonal self-energy diagram mediated by a couple of Higgs and lepton doublets. At this stage the total lepton number, defined as the sum of the lepton asymmetries in the active and sterile flavours, is still vanishing. However, since the Yukawa couplings are in general complex numbers, individual lepton asymmetries are generated in the different flavours, and in general a net lepton asymmetry appears in the active sector, which is equal in magnitude and opposite in sign with respect to the asymmetry created in the sterile sector. Since the SM sphalerons only couple to the active leptons, they convert the lepton asymmetry in the active flavours (and only this asymmetry) into a net baryon asymmetry.

\section{Naturalness argument}
The flavoured leptogenesis scenario is based on the assumption that the Yukawa couplings for the new sterile fermions are sufficiently small such that these new states do not equilibrate before the electroweak phase transition. Since the final lepton asymmetry is proportional (among other parameters) to the abundance of sterile fermions, it will be in general suppressed in the flavoured leptogenesis scenario. There are two possible configurations of parameters that can boost the asymmetry production in order to account for the observed value: the first one is to have a hierarchical structure in the Yukawa couplings, such that $|Y_{\alpha I}|\ll Y_{eq}$ for certain flavours $\alpha$, while $|Y_{\beta I}|\gg Y_{eq}$ for the others. In this way the large Yukawas ensure a sizeable production of sterile fermions, while the small Yukawas protect the asymmetry generated in the flavours $\beta$ from washout.
The second possibility is to have at least one pair of sterile fermions that are strongly degenerate in mass: the degeneracy enhances the oscillation rate among different flavours, allowing for the production of sizeable asymmetries.

Although both these particular configurations of parameters are legitimate, they appear to be fine-tuned solutions if they are not justified by symmetry arguments.
In the present work we focus on the naturalness of the second scenario, proposing the hypothesis of having an approximate lepton number symmetry as a key to achieve a low scale (flavoured) leptogenesis~\cite{Abada:2015rta}. Neutrino mass generation mechanisms based on an approximate lepton number symmetry are a well motivated scenario, in which the smallness of the neutrino masses (when compared to the electroweak scale) is related to the smallness of the new lepton  number violating parameters, which are natural in the sense of 't Hooft, since the Lagrangian acquires a new symmetry when they are set to zero. Therefore neutrino masses are stable against radiative corrections. Moreover the assumption of an approximate lepton number conservation is phenomenologically justified from the fact that no violation of lepton number has been observed so far.
Our hypothesis is driven by the observation that, in the seesaw mechanisms characterised by an approximate lepton number symmetry, the heavy fermionic singlets naturally couple to form pseudo-Dirac pairs strongly degenerate in mass. Indeed in the limit of exact symmetry the heavy Majorana states must preserve total lepton number, i.e. they must either form a Dirac particle or decouple, while neutrino masses must vanish.

The minimal setup in order to implement this idea is to extend the SM by adding a pair of sterile fermions, $N_{1,2}$, with opposite lepton number, $L=\pm 1$. In a simplified setup with only one active flavour, the lepton number conserving part of the neutrino mass matrix reads, in the basis $(\nu_L, {N_1}^c, {N_2}^c)$,
\bee\label{eq:M0_toy}
\mathcal{M}_0 &=& \left(\begin{array}{ccc} 0 & y v & 0\\
y v & 0 & \Lambda \\
0 & \Lambda & 0
\end{array}\right),
\eee  
where $y$ is a dimensionless Yukawa coupling and $\Lambda$ is a parameter with dimensions of mass. For the sake of presentation we assume all parameters to be real in this toy model. The ``lepton number conserving'' mass spectrum resulting from the diagonalization of this mass matrix is composed by a massless state $m_\nu \equiv M_1= 0$, and two Majorana massive states that combine to form a Dirac massive state, $M_2 = M_3 = \sqrt{\Lambda^2 + v^2 y^2}$.

\section{Some minimal mechanisms}
In order to account for both the observation of massive neutrinos and to achieve a successful leptogenesis it is necessary to perturb the lepton number conserving mass matrix (\ref{eq:M0_toy}) by adding small lepton number violating entries. Barring a non-zero element in the $(1,1)$ entry of the matrix in eq.~(\ref{eq:M0_toy}), which corresponds to a gauge violating Majorana mass term for left-handed neutrinos and requires a non minimal extension of the SM (for example the addition of an Higgs isospin triplet), there are 3 possibilities to do it, resulting into the well-known patterns of Inverse Seesaw (ISS), Linear Seesaw (LSS) and Extended Seesaw (ESS), 
\begin{equation}
\begin{array}{lcccr}
\Delta M_{ISS}=\left(\begin{array}{ccc}
 0 & 0 & 0\\
0 & 0 & 0\\
0 & 0 & \xi \Lambda
\end{array}\right),& \hspace{-0.03\textwidth} 
&
\Delta M_{LSS}=\left( \begin{array}{ccc} 0 & 0 & \epsilon y v\\
0 & 0 & 0\\
\epsilon\, y v & 0 & 0 
\end{array}\right),& \hspace{-0.03\textwidth} 
&
\Delta M_{ESS}=\left(\begin{array}{ccc} 0 & 0 & 0\\
0 & \xi' \Lambda & 0\\
0 & 0 & 0 
\end{array}\right),
\end{array}
\label{eq_DMs}
\end{equation}
where $\xi, \epsilon, \xi'$ are small $(<1)$ dimensionless parameters. The values for the active neutrino mass scale $m_\nu$, and for the mass squared difference among the heavy states $\Delta M_{32}^2 = M_3^2-M_2^2$, are reported in Table~\ref{tab:toymin} for each mechanism, at the leading order in the $\xi, \epsilon, \xi'$ parameters.

\begin{table}[htb]
\begin{center}
\caption{Active neutrino mass scale and mass squared splitting in the heavy pseudo-Dirac pair for each one of the minimal perturbations (\ref{eq_DMs}), to the lowest order in the small lepton number violating parameters $\xi,\epsilon,\xi'$. $f(x)$ is a loop function such that $f(x)\ll 1$ for $x<1$.}
\label{tab:toymin}
\begin{tabular}{|c|c|c|c|}
\hline 
& ISS & LSS & ESS\\
\hline
$m_\nu$ & $\xi y^2 \frac{v^2}{\Lambda}$ &$2 \epsilon y^2 \frac{v^2}{\Lambda}$ & $\xi' y^2 \frac{v^2}{\Lambda} f\left(\frac{\Lambda^2}{M_W^2}\right)$ \\ 
\hline 
$\Delta M^2_{32}$ & $2 \xi \Lambda^2$& $4 \epsilon v^2 y^2$ & $2 \xi' \Lambda^2$\\
\hline
\end{tabular}
\end{center}
\end{table}

By imposing the requirements of having a sufficiently large neutrino mass scale, $m_\nu \gtrsim \sqrt{\Delta m_\textrm{atm}^2}\simeq 5 \times 10^{-2}$ eV, together with sufficiently small Yukawa couplings, $y < \sqrt{2} \times 10^{-7}$, we conclude that the ISS mechanism predicts a too large mass splitting in the heavy pseudo-Dirac pair in order to account for a successful leptogenesis~\cite{Abada:2015rta}. On the contrary, the mass splitting in the LSS  depends on the Higgs vev $v$, which vanishes in the unbroken phase. The ESS suffers from the same issues of the ISS, but they are still more accentuated. From this analysis we conclude that the minimal extension of the SM, based on an approximate lepton number conservation and able to account for both the observed neutrino masses and for a successful GeV-scale leptogenesis, is the addition of a pair of sterile fermions with opposite lepton number, considering both an ISS and a LSS-like perturbations.

In the realistic scenario with 3 active flavours the mass matrix for this model is given by~\cite{Gavela:2009cd}
\begin{equation}\label{eq_Mpertexp}
 \mathcal{M} =  \left( \begin{array}{ccc}
 \mathbf{0} & \textbf{Y} v &\epsilon \textbf{Y}' v\\
 \textbf{Y}^T v & 0 & \Lambda \\
 \epsilon{\textbf{Y}'}^T v & \Lambda & \xi \Lambda 
 \end{array}
\right),
\end{equation}
where $\textbf{Y},\textbf{Y}'$ are now 3-dimensional vectors, whose entries are assumed to be of the same order of magnitude. Notice that the ordering of the second and third column/row  of eqs.~(\ref{eq:M0_toy}, \ref{eq_Mpertexp}) arises from the assignment $L=+1$ and $-1$, for $N_1$ and $N_2$, respectively. Choosing $\epsilon > 1$ and $|\textbf{Y}| \simeq |\textbf{Y}'|$ correspondingly smaller implies inverting this assignment. Thus very large values of $\epsilon \gg 1$ also correspond to an approximate lepton number conservation, and there is an approximate symmetry under $\epsilon \rightarrow 1/\epsilon$, which becomes exact when $\xi \rightarrow 0$.

\begin{figure}[htb]
 \begin{tabular}{cc}
\includegraphics[width=0.45\textwidth]{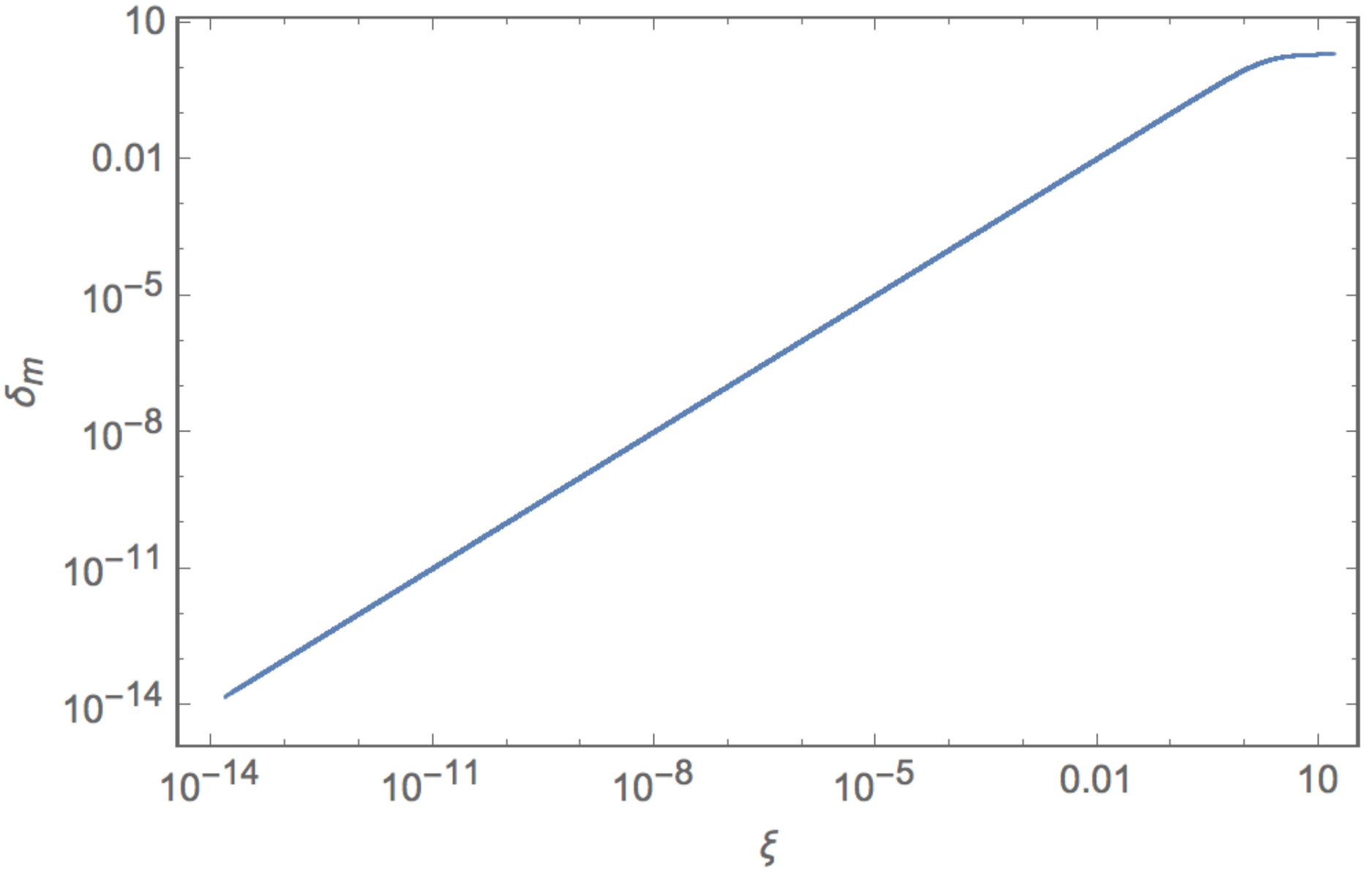}
\hspace*{2mm}&\hspace*{2mm}
\includegraphics[width=0.45\textwidth]{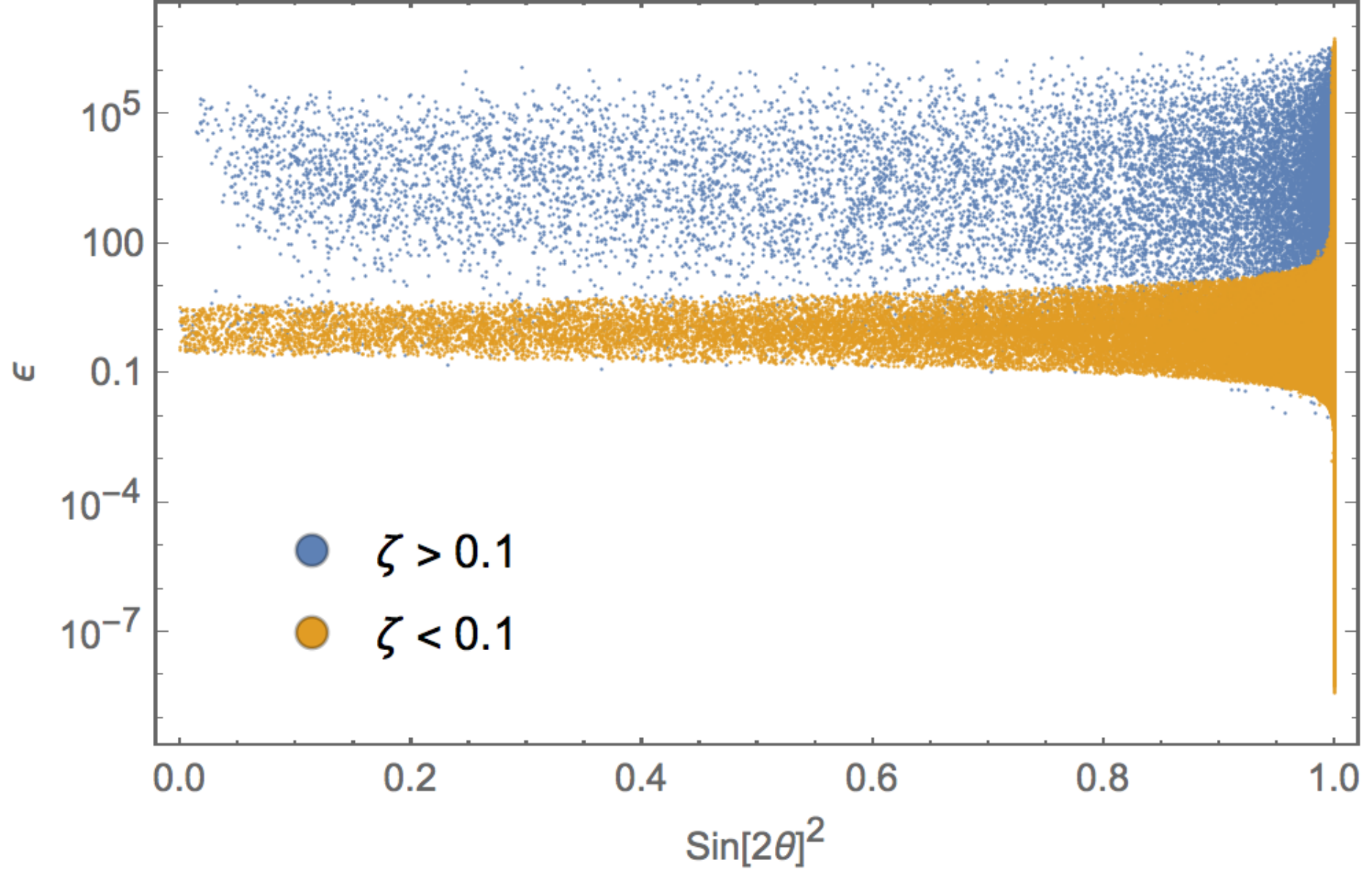}
\end{tabular}
\caption{Correlation between the lepton number violating parameters and the quantities governing the oscillation dynamics in the heavy pseudo-Dirac pair. \emph{Left panel}: relative mass splitting $\delta_m$ in the pseudo-Dirac pair composed by two states of masses $M_4$ and $M_5$,  $\delta_m \equiv 2 (M_5-M_4)/(M_5+M_4)$, as a function of the ``ISS''-like perturbation $\xi$. \emph{Right panel}: mixing angle $\sin(2 \theta)^2$ as a function of the ``LSS''-like perturbation $\epsilon$; blue (orange) points refer to solutions with $\xi>0.1$ ($\xi<0.1$).}
\label{fig:param_osc}
\end{figure}

It is interesting to notice that the small lepton number violating parameters $\epsilon$ and $\xi$ are directly related to the dynamics of the oscillations in the pseudo-Dirac pair: the regime $\epsilon \ll 1$ implies a large mixing angle, while $\xi \ll 1$ implies a small small splitting, cf. Fig.~\ref{fig:param_osc}. Thus, in a scenario with an approximate lepton number symmetry the flavoured leptogenesis is naturally enhanced.

\section{Viable parameter space and testability}
In the weak-washout regime it is possible to derive an analytic expression for the final baryon asymmetry in the considered model~\cite{Abada:2015rta}:
\begin{equation}
\label{eq:baryo_analytical}
Y_{\Delta B}=\frac{n_{\Delta B}}{s}= \frac{945}{2528} \frac{\, 2^{2/3}}{  \,\,  3^{1/3} \, \pi^{5/2}  \,  \Gamma(5/6)} \frac{1}{g_s}\sin^3 \phi \, \frac{M_0}{T_{\rm W}} \frac{M_0^{4/3}}{ \left(\Delta m^2\right)^{2/3}} \, Tr\left[ F^\dagger \delta F\right] \ ,
\end{equation}
where $F_{\alpha j}  = Y_{\alpha I} U_{I j}$ are the Yukawa couplings in the mass basis, $U$ is the lepton mixing matrix and the indices run over $\alpha = e,\mu,\tau$ (active flavours), $I=1,2$ (sterile flavours) and $j=1,\dots, 5$ (mass eigenstates). 
 $\Delta m^2=M_5^2-M_4^2$ is the mass squared difference in the pseudo-Dirac pair, $g_s$ represents the degrees of freedom in the thermal bath at $T = T_{W}\simeq 140$ GeV, $M_0 \approx 7 \times 10^{17}\,\mbox{GeV}$, $\sin\phi \sim 0.012$ and $\delta = \textrm{diag}(\delta_\alpha)$ is defined as:
\begin{equation}\label{eq:deltaCP}
\delta_{\alpha}=\sum_{i>j} \textrm{Im}\left[F_{\alpha i} \left(F^{\dagger} F\right)_{ij} F^{\dagger}_{j\alpha}\right]\ .
\end{equation}

\begin{figure}[htb]
\begin{center}
\subfloat{\includegraphics[width=6.0 cm]{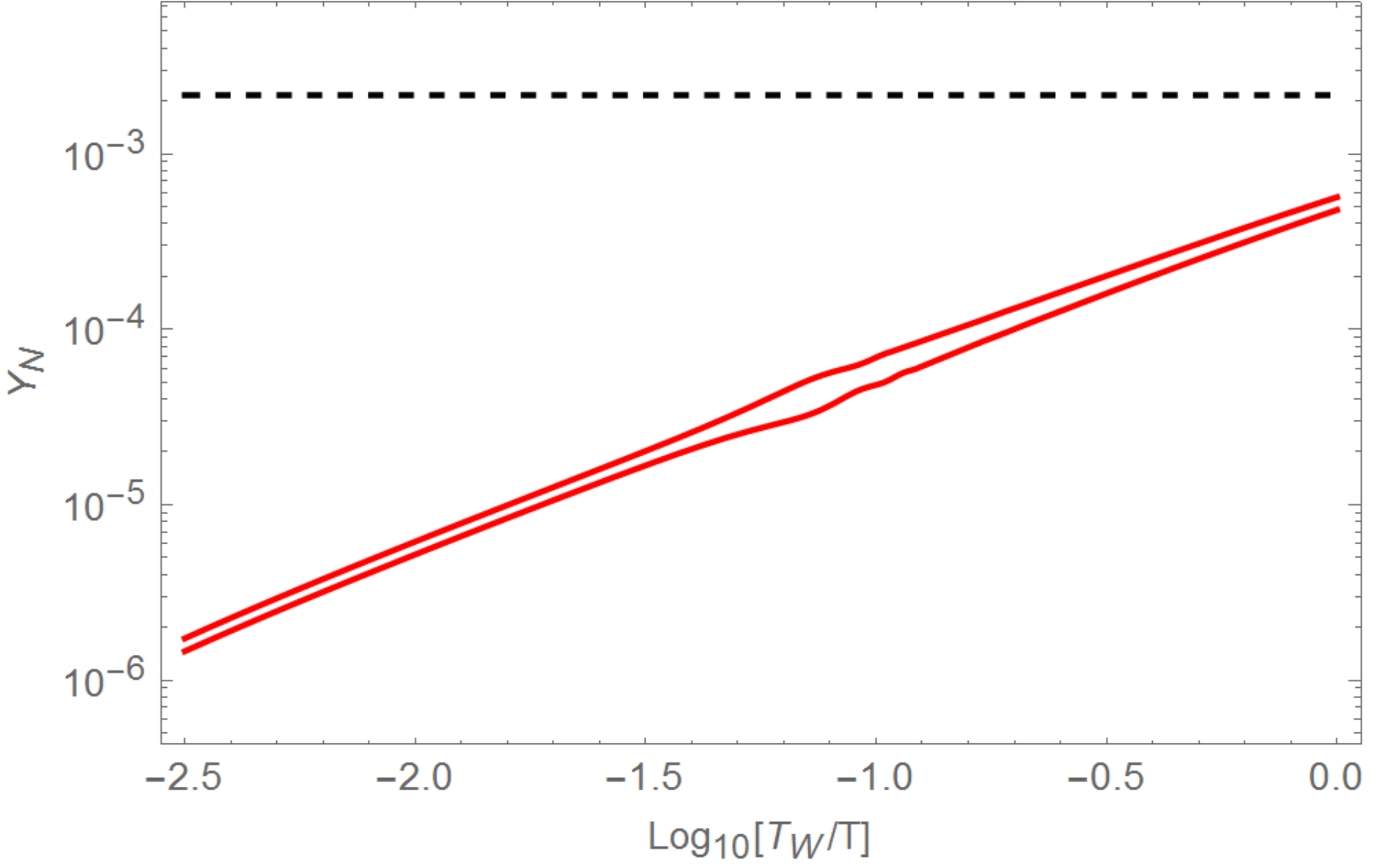}}
\subfloat{\includegraphics[width=6.0 cm]{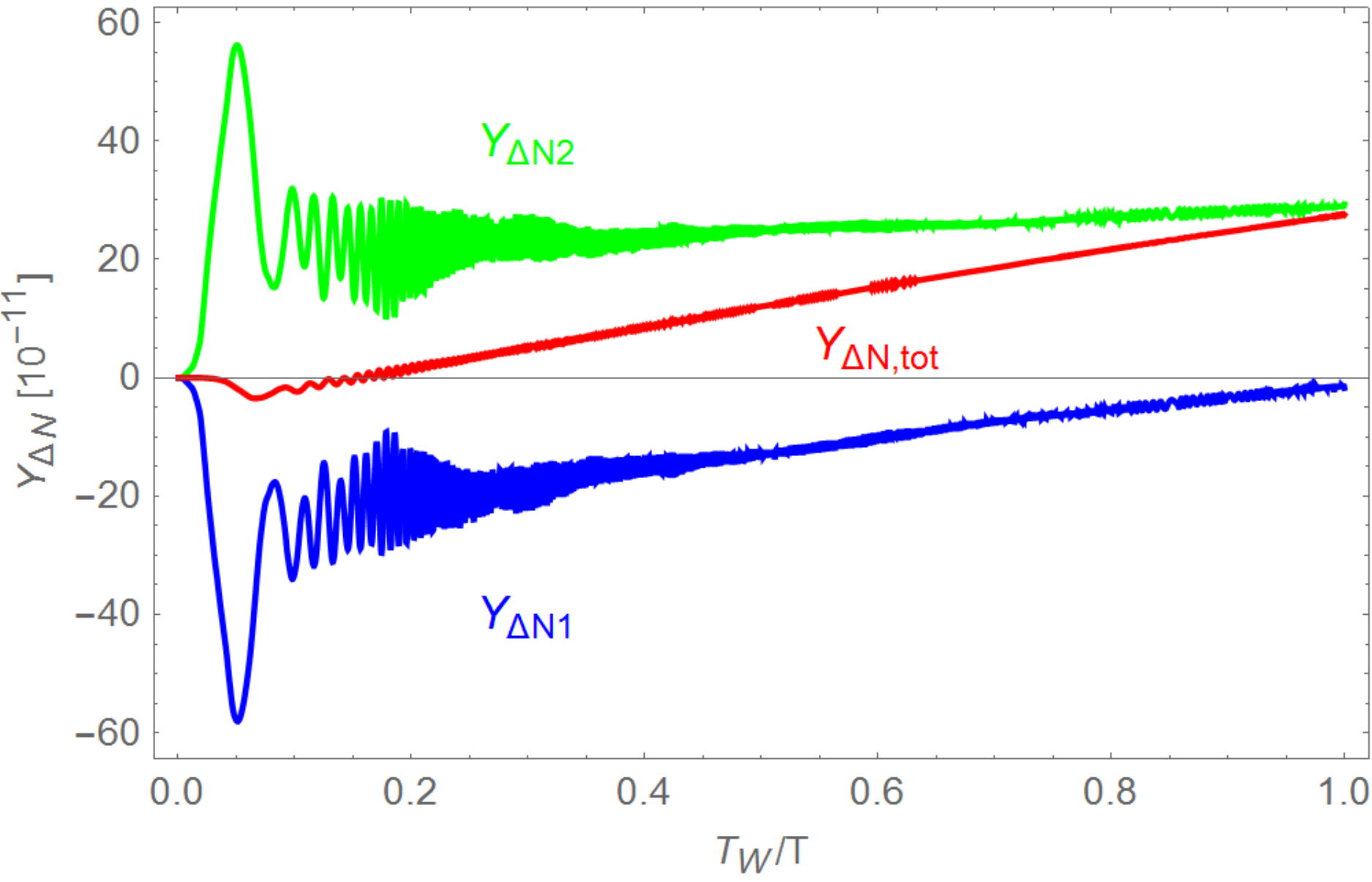}}\\
\subfloat{\includegraphics[width=6.5 cm]{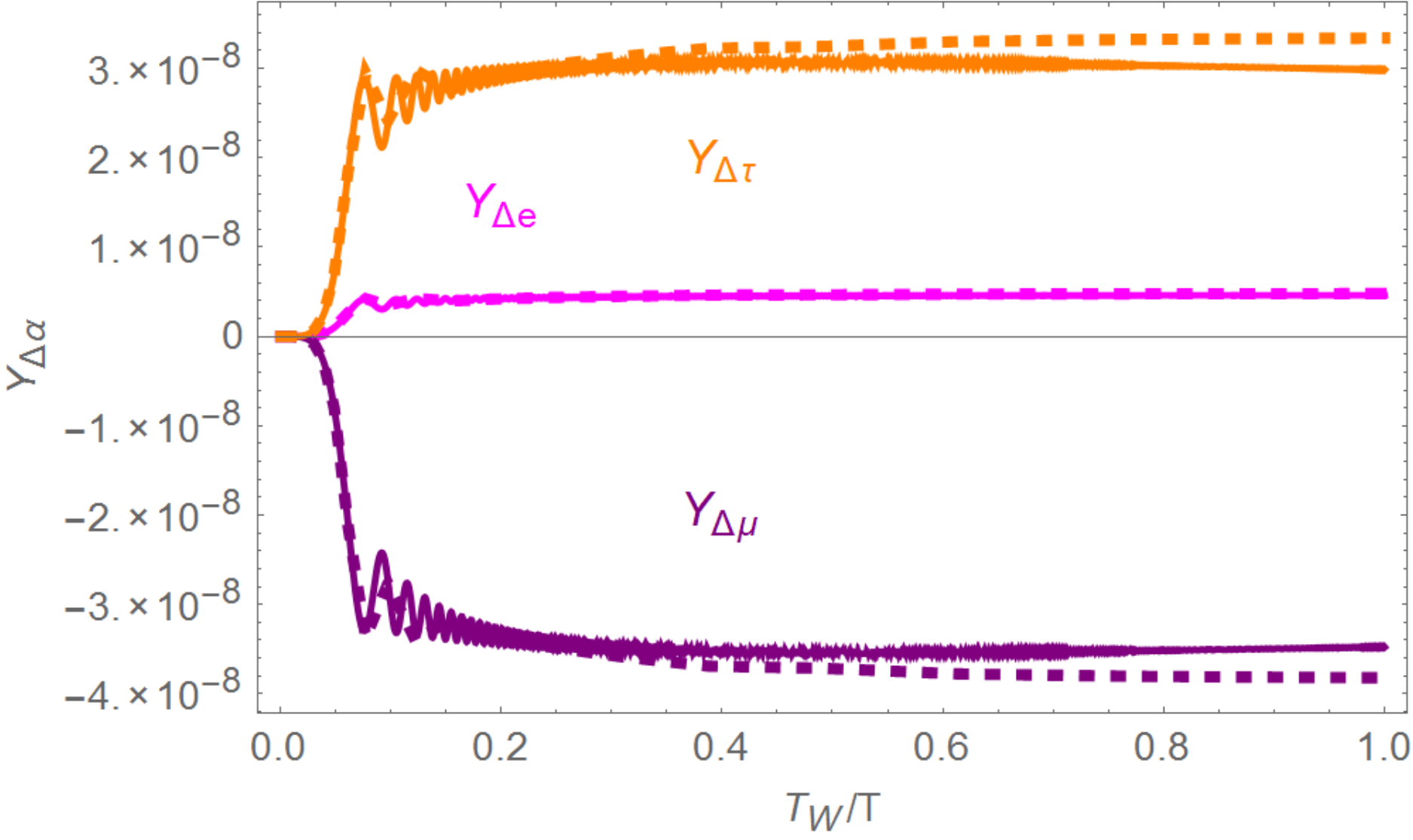}}
\subfloat{\includegraphics[width=6.5 cm]{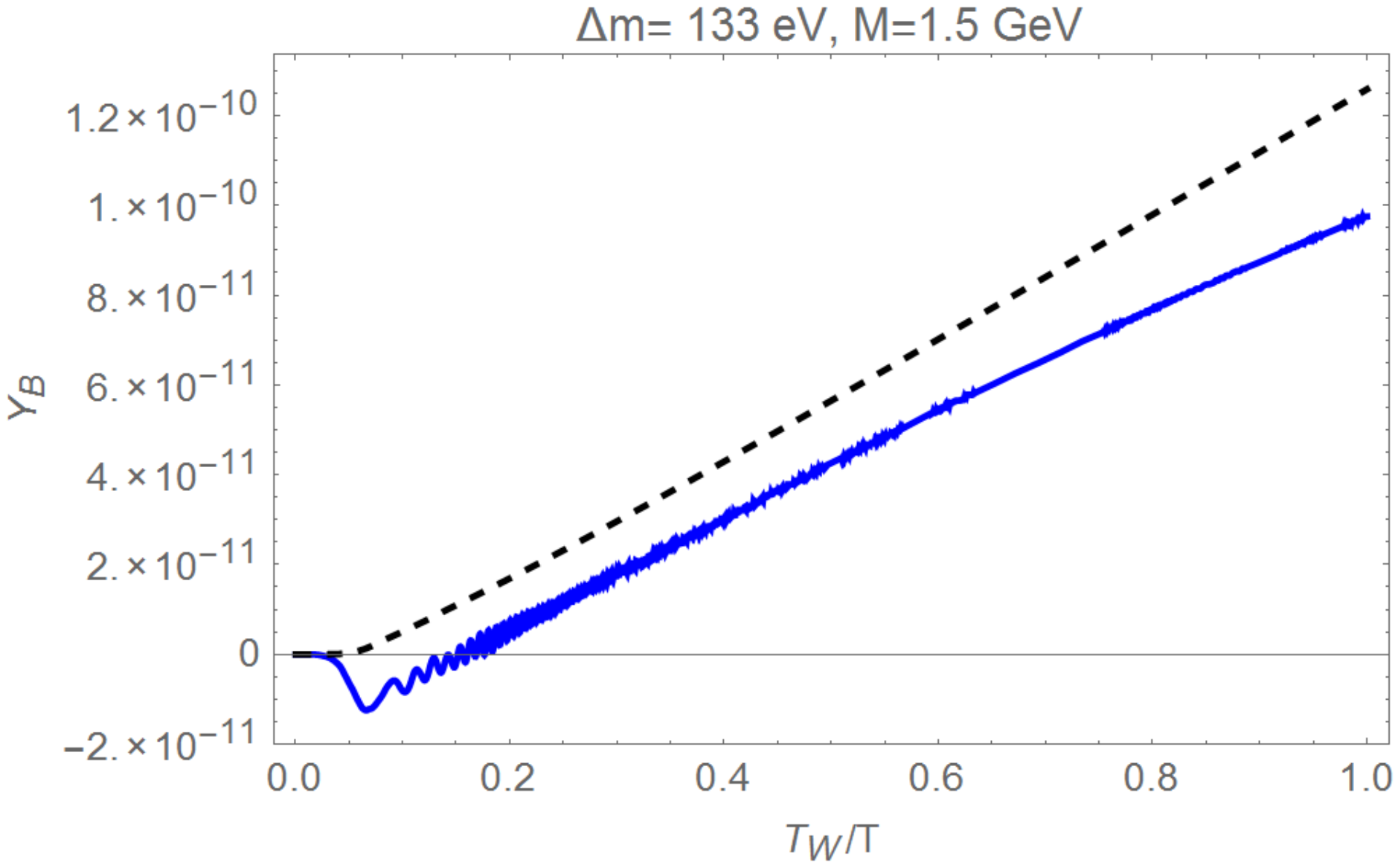}}
\caption{Numerical resolution of a benchmark point lying in the weak-washout regime, and comparison with the derived analytical expressions. \emph{Left top panel}: Evolution of the abundance of the heavy fermions (red solid lines) compared with their equilibrium value (dashed black line). \emph{Right top panel}: Evolution of the total (red line) and individual (blue and green lines) asymmetries in the two sterile flavours as a function of the temperature. 
\emph{Left bottom panel}: Evolution of the asymmetries in the active flavours according the numerical solution of the Boltzmann equations (solid lines) and the analytical estimate (dashed lines). 
\emph{Right bottom panel}: Evolution of the baryon yield with temperature (blue line) compared with its analytical determination (dashed black line).
}
\label{fig:bench_natural}
\end{center}
\end{figure}

This analytical expression has been validated and complemented by means of the numerical resolution of the system of Boltzmann equations for a set of selected benchmark points~\cite{Abada:2015rta}, one of which is reported in Fig.~\ref{fig:bench_natural}. We found a good agreement between the analytic and numerical solutions as long as the condition $|F_{\alpha j}| \lesssim \sqrt{2} \times 10^{-7}$ is satisfied. For larger Yukawa couplings, washout effects become important and the expression~(\ref{eq:baryo_analytical}) overestimates the final asymmetry.

\begin{figure}[htb]
 \begin{tabular}{cc}
\includegraphics[width=0.45\textwidth]{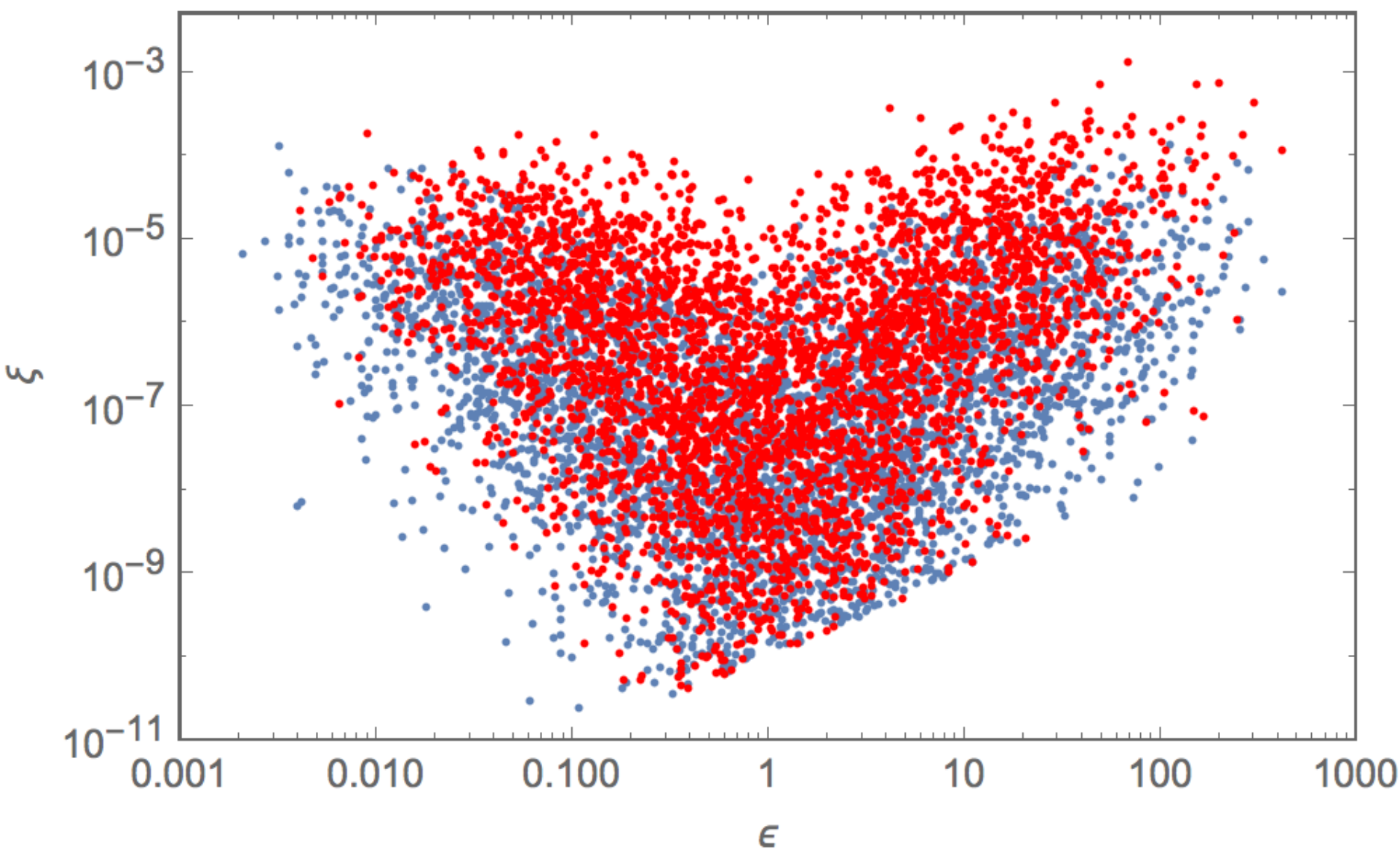}
\hspace*{2mm}&\hspace*{2mm}
\includegraphics[width=0.45\textwidth]{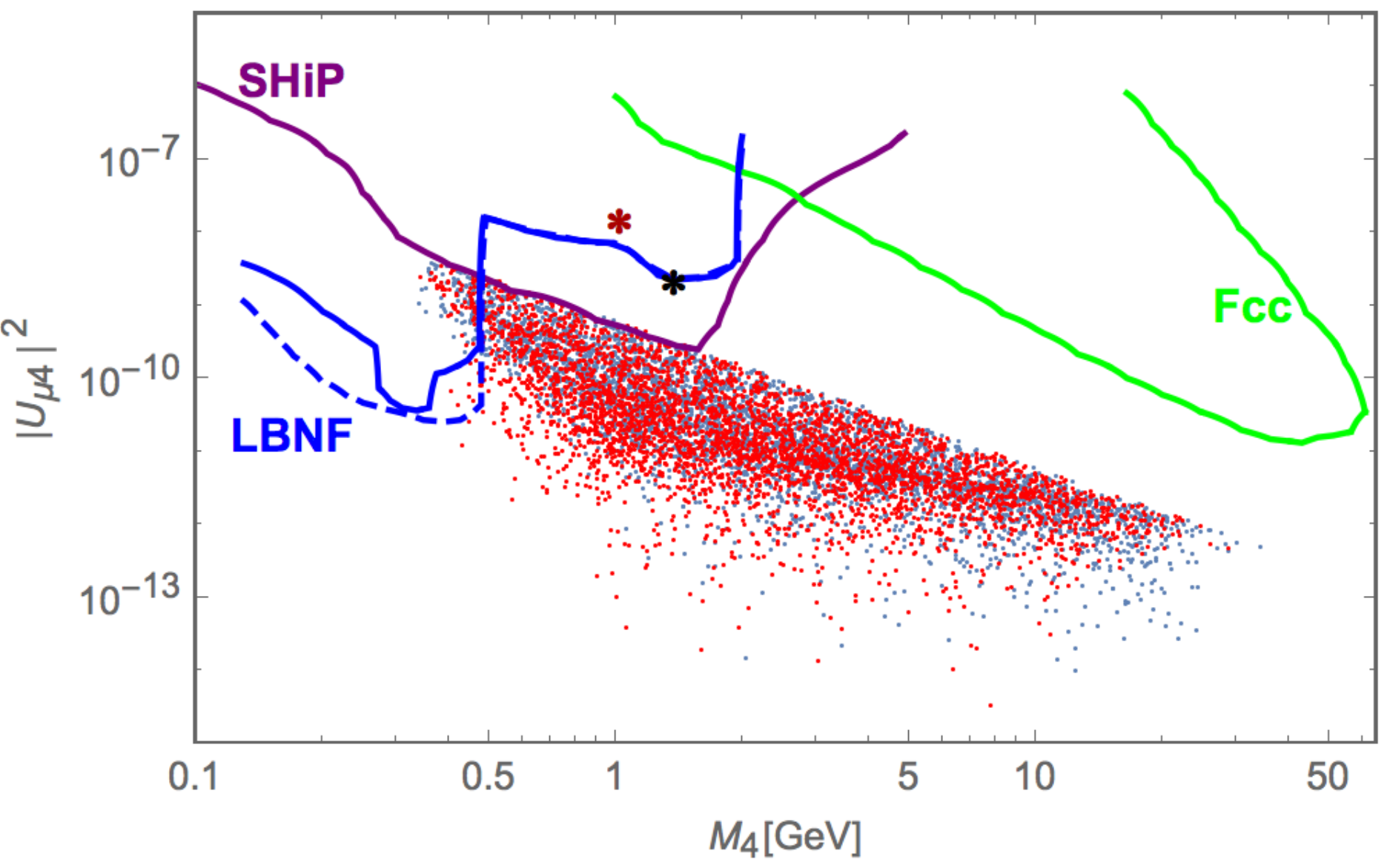}
\end{tabular}
\caption{Viable solutions accounting for both neutrino parameters and leptogenesis in the weak-washout regime. Blue (red) points refer to a normal (inverse) hierarchy for the light neutrino mass spectrum. \emph{Left panel}: viable parameter space for the lepton number violating parameters $\xi,\epsilon$. \emph{Right panel}: viable solutions in the plane ($|U_{\mu 4}|^2$, $M_4$), where $U_{\mu 4}$ is the active-sterile mixing in the $\mu$ flavour for the lightest sterile fermion with mass $M_4$; the expected sensitivity of some planned future experiments is reported for comparison. The asterisks refer to two viable solutions in the strong-washout regime.}
\label{fig:param_pert}
\end{figure}

\begin{figure}[h!]
\begin{center}
\subfloat{\includegraphics[width=6.2 cm]{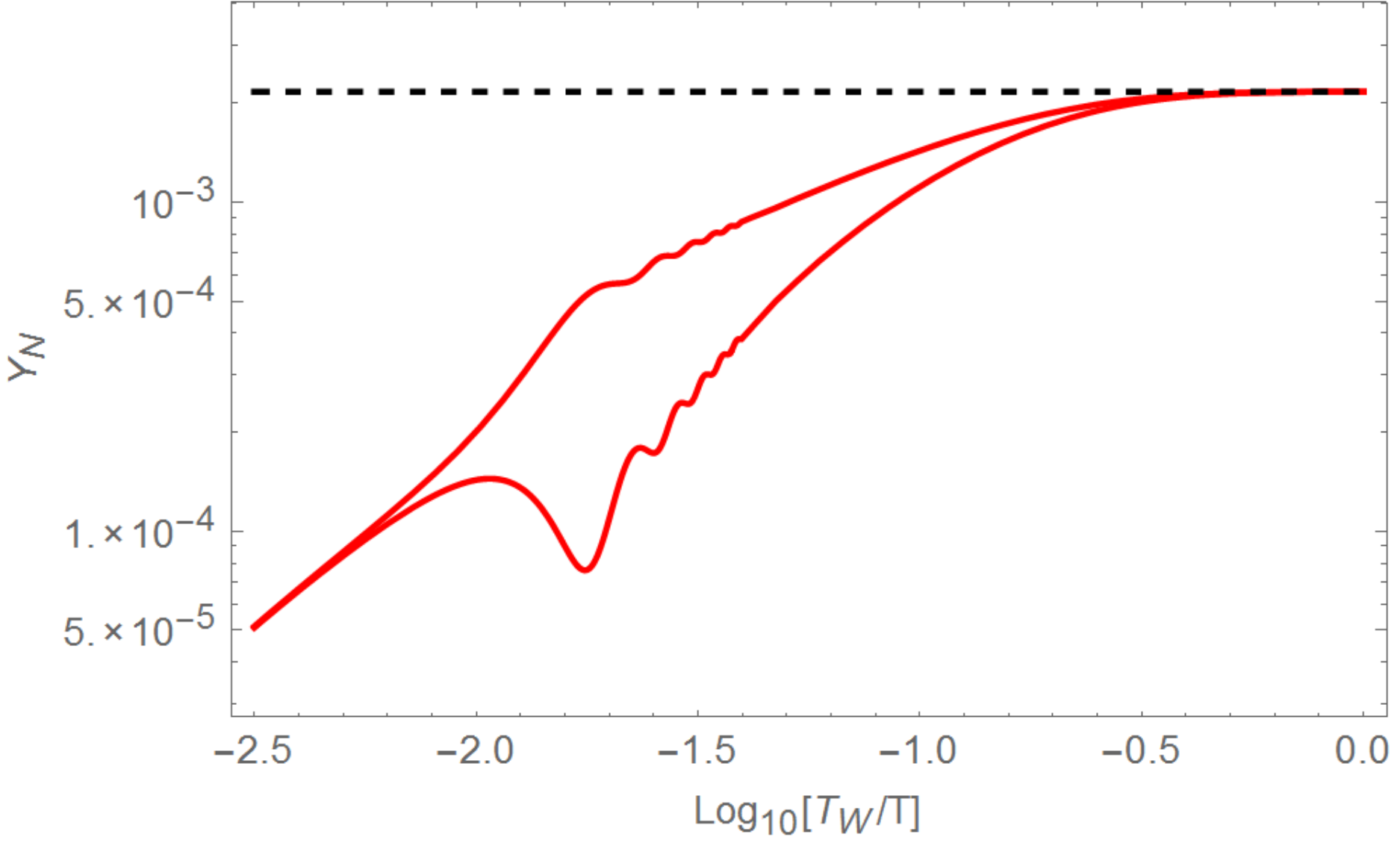}}
\subfloat{\includegraphics[width=6.0 cm]{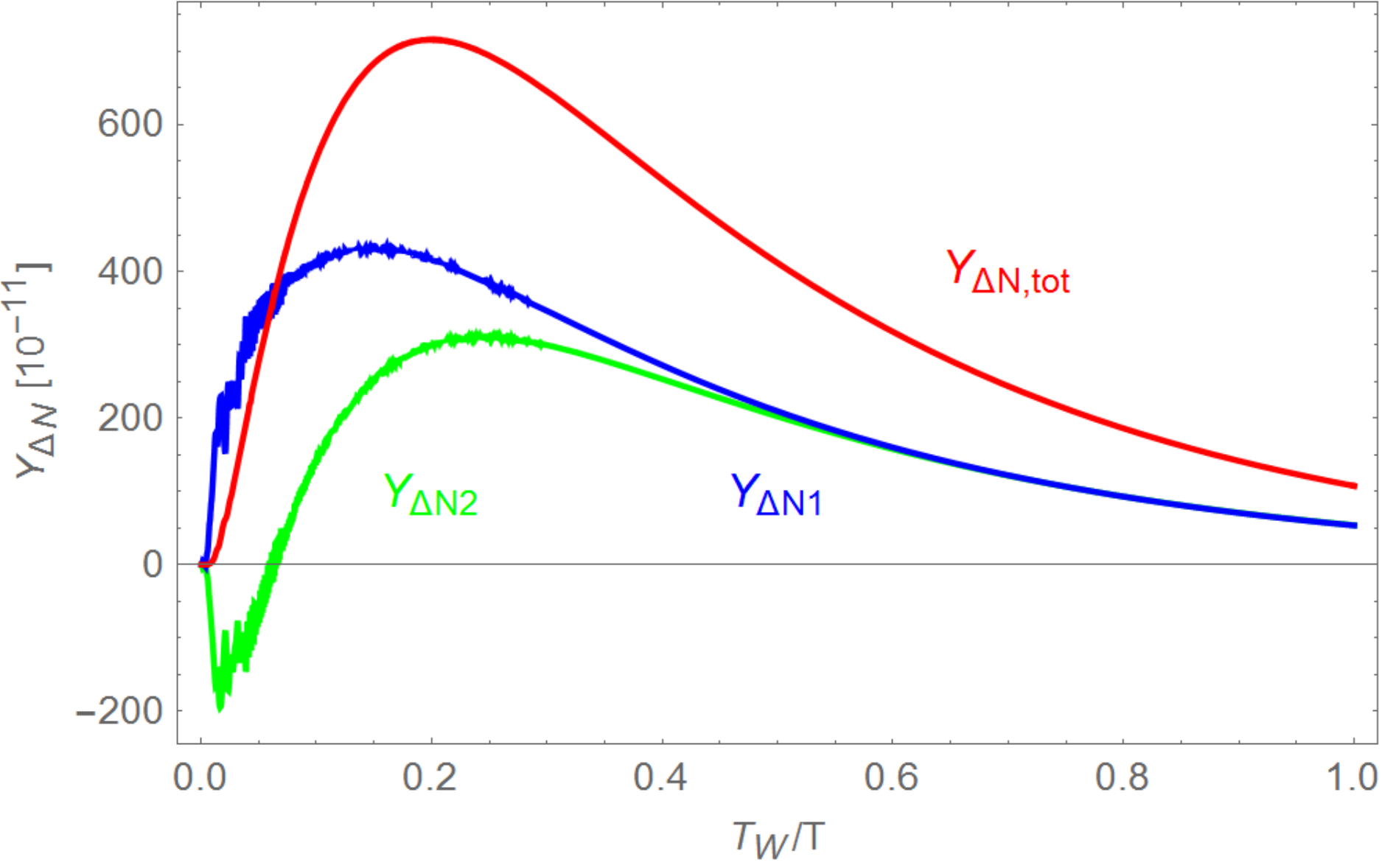}}\\
\subfloat{\includegraphics[width=6.5 cm]{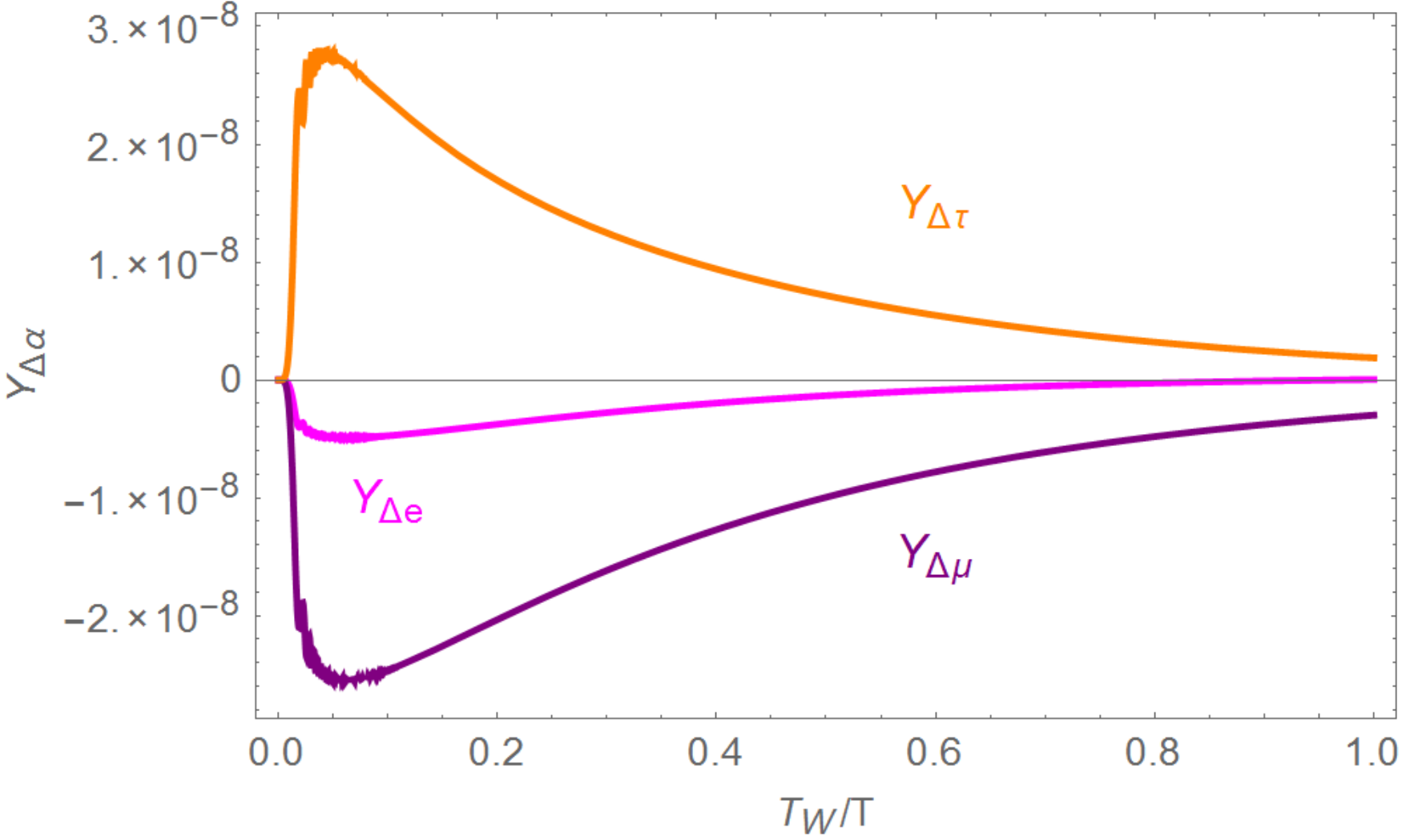}}
\subfloat{\includegraphics[width=6.5 cm]{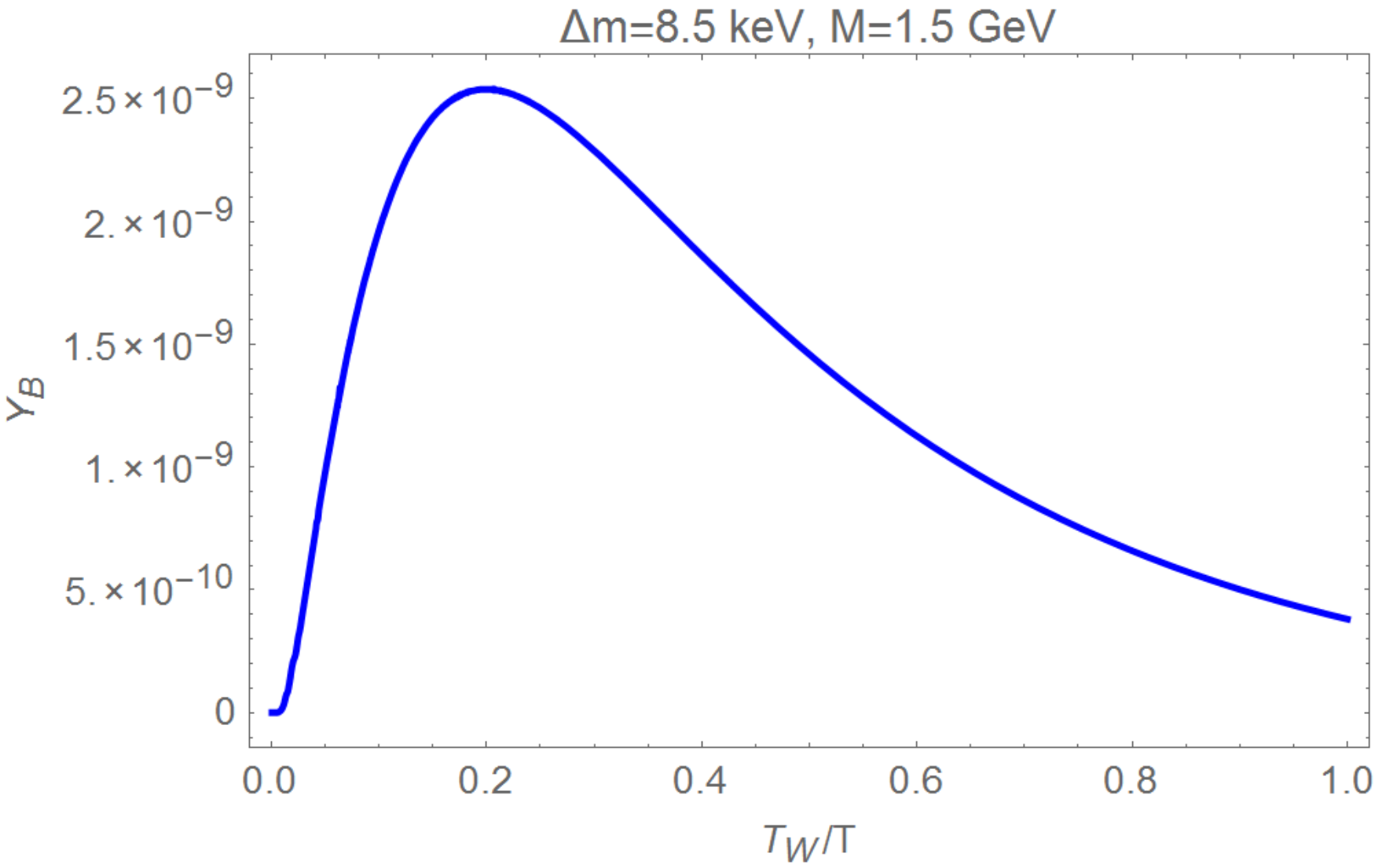}}
\caption{As in Fig.~\ref{fig:bench_natural} but for a benchmark point lying in the strong-washout regime. As it it evident the heavy fermions equilibrate at late times, and there is a consequent depletion of the produced asymmetry. However this depletion is not complete, and a residual baryon asymmetry of the order of the observed one survives at the sphaleron freeze-out temperature.}
\label{fig:benchhy2}
\end{center}
\end{figure}

We use the relation~(\ref{eq:baryo_analytical}) to perform a scan of the parameter space of the model in the weak-washout regime, in order to identify the solutions that can account for both the neutrino data and the observed baryon asymmetry. The results are reported in Fig.~\ref{fig:param_pert}: in the left panel the allowed values for the lepton number violating parameters $\epsilon$ and $\xi$ are reported. In the right panel we report the values for the active-sterile mixing of the pseudo-Dirac pair in the $\mu$ flavour as a function of its mass, together with the expected sensitivity of the future experimental facilities SHiP, LBNF/DUNE and Fcc~\cite{Adams:2013qkq,Alekhin:2015byh}. We see that having limited the analysis to the weak-washout regime sets an upper bound on the allowed active-sterile mixing values (as a consequence of the upper bound on the Yukawa couplings), and the allowed solutions lie outside the experimental reach, apart from a limited region at small masses that could be probed by LBNF. This result however does not imply that the model is not testable at future facilities, since we have limited the analysis to the subregion of the parameter space where eq.~(\ref{eq:baryo_analytical}) is valid. Outside this region we must rely on the numerical resolution of the system, which is computationally very demanding and prevents us from performing a complete scan. We opt for numerically solving a set of benchmark points in the strong-washout regime~\cite{Abada:2015rta}, one example of which is reported in Fig.~\ref{fig:benchhy2}, and we report in the right panel of Fig.~\ref{fig:param_pert}  two solutions that give the correct baryon asymmetry (asterisks). It is evident from this analysis that testable solutions in the reach of future facilities exist, and the viable parameter space extends above the weak-washout region scanned in Fig.~\ref{fig:param_pert}. A subsequent study aiming at exploring the entire parameter space is in progress.

\section{Conclusion}
In this work~\cite{Abada:2015rta} we proposed the hypothesis of having an approximate lepton number symmetry as a way to achieve successful leptogenesis in low-scale neutrino mass generation mechanisms, characterised by the appearance of new physics at the GeV scale. We studied the minimal extension of the SM in order to implement the idea, which results in the addition of a pair of sterile fermions with opposite lepton number, and considering both ISS and  LSS-like lepton number violating operators. We provide an analytic expression for the final asymmetry which is valid in the weak-washout regime, and perform a scan of the solutions in this region of the parameter space. We also solve numerically the system of Boltzmann equations for a set of benchmark points, both in the weak and  strong-washout regions. We conclude that solutions exist in a sizeable portion of the weak-washout region, but these points are outside the sensitivity of future experiments. We also find that viable solutions exist in the strong-washout region, and that these points are testable by future facilities.

\section*{Acknowledgments}
M.L. is supported  by the Belgian Federal Science Policy Office through the Interuniversity Attraction Pole P7/37. He acknowledges financial support from the 51st Rencontres de Moriond conference.

\end{document}